\documentclass[14pt,a4paper]{article}

\usepackage[english]{babel}
\usepackage{amsmath,amssymb,amsfonts,amsthm}
\usepackage[mathscr]{eucal}
\usepackage[all]{xy}
\usepackage{hyperref}
\usepackage{setspace}
\usepackage{upgreek}
\usepackage{hyperref}

\usepackage{times}
\usepackage{color}

\usepackage{indentfirst}

\begin{document}

\title{\textbf{Multi-Hamiltonian formulations and stability  \\ of  higher-derivative extensions  of $3d$
Chern-Simons }}
\author {V.~A.~Abakumova\footnote{abakumova@phys.tsu.ru}, \ D.~S.~Kaparulin\footnote{dsc@phys.tsu.ru}, \ and S.~L.~Lyakhovich\footnote{sll@phys.tsu.ru}}
\date{\footnotesize\textit{Physics Faculty, Tomsk State University, Tomsk 634050, Russia }}

\maketitle

\begin{abstract}
\noindent Most general third-order  $3d$ linear gauge vector field
theory is considered. The field equations involve, besides the mass,
two dimensionless constant parameters. The theory  admits
two-parameter series of conserved tensors with the canonical
energy-momentum being a particular representative of the series. For
a certain range of the model parameters, the series of conserved
tensors include bounded quantities. This makes the dynamics
classically stable, though the canonical energy is unbounded in all
the instances. The free third-order equations are shown to admit
constrained multi-Hamiltonian form with the zero-zero components of
conserved tensors playing the roles of corresponding Hamiltonians.
The series of Hamiltonians includes the canonical Ostrogradski's
one, which is unbounded.  The Hamiltonian formulations with
different Hamiltonians are not connected by canonical
transformations. This means, the theory admits inequivalent
quantizations at the free level. Covariant interactions are included
with spinor fields such that the higher-derivative dynamics remains
stable at interacting level if the  bounded conserved quantity
exists in the free theory. In the first-order formalism, the
interacting theory remains Hamiltonian and therefore it admits
quantization, though the vertices are not necessarily Lagrangian in
the third-order field equations.
\end{abstract}

\section{Introduction}

Classical dynamics and quantization of various higher-derivative
models are discussed once and again for decades. Among most
frequently studied specific models we can mention  Pais-Uhlenbeck
(PU) oscillator \cite{PU}, Podolsky and Lee-Wick electrodynamics
\cite{Pod}, \cite{LW1}, \cite{LW2}, higher-derivative extensions of
the Chern-Simons model \cite{Deser}, higher-derivative Yang-Mills
models \cite{SlavFad}, conformal gravity \cite{Conf}, various
higher-derivative higher-spin fields theories \cite{HS1},
\cite{HS2}, \cite{HS3}, modified theories of gravity \cite{Grav1},
including critical gravity \cite{Crit}.  The higher-derivative
models often reveal remarkable various properties comparing to the
counterparts without higher derivatives. In particular, the
inclusion of the higher-derivatives improves the convergency in the
field theory both at classical and quantum level in many models.
Also the conformal symmetry often requires inclusion of higher
derivatives in the field equations.

The higher-derivative dynamics are also notorious for the classical
and quantum instability. The key point, where the problem can be
immediately seen is that the canonical energy is unbounded for
general higher-derivative Lagrangian systems. Several exceptions are
known \cite{And}, \cite{Chen}, \cite{Berg1}, \cite{Berg2},
\cite{fR1}, \cite{fR2} of the higher-derivative models such that
have bounded canonical energy. In all these cases, the energy is
bounded on shell because of strong constraints among the field
equations. At the quantum level, the instability reveals itself by
the ghost poles in the propagator and it is related to the problem
of unbounded spectrum of energy. In its turn,  the unbounded energy
spectrum results from the fact that canonical Hamiltonian, being the
phase space equivalent of the canonical Noether's energy, is
unbounded due to the higher derivatives.

In the work \cite{KLS14}, it was noticed that the broad class of
higher-derivative models are stable at classical level because they
admit conserved tensors with bounded $00$-component. The bounded
conserved quantity turns out different from the canonical energy
which can be unbounded for the same dynamics. Furthermore, these
models admit non-canonical Lagrange anchors\footnote{The concept of
Lagrange anchor was introduced in ref. \cite{KazLS} to covariantly
quantize not necessarily Lagrangian field theories. Later, it was
established that the Lagrange anchor maps global symmetries to
conserved currents \cite{KLS10}. The canonical Lagrange anchor,
being an identity map, is always admitted by the Lagrangian
equations, and it identifies a symmetry with the characteristics of
conserved quantity. This can be understood as the Noether theorem in
a different wording. The non-canonical Lagrange anchor also connects
the symmetries with characteristics of conserved currents, though it
is not an identity map. Every Lagrange anchor connects any conserved
current to a certain symmetry. Once the dynamical system, be it
Lagrangian or not, admits different Lagrange anchors, the same
symmetry can be connected with different conserved quantities.}. The
class of higher-derivative systems considered in ref. \cite{KLS14}
covers a variety of well known higher-derivative models, including
PU oscillator and Podolsky electrodynamics. In ref. \cite{KKL}, this
class is further generalized, and it covers also extended
Maxwell-Chern-Simons models. In all the examples of
higher-derivative models considered in \cite{KLS14}, \cite{KKL}, the
conserved tensor with bounded $00$-component turns out connected
with space-time translations by the non-canonical Lagrange anchor.
In this sense, the bounded conserved tensor can be interpreted as a
non-canonical energy-momentum. The conservation law makes the theory
stable at classical level irrespectively to interpretation of the
conserved quantity. Also notice that all the considered examples
\cite{KLS14}, \cite{KKL} of stable higher-derivative models admit
the interactions such that do not spoil classical stability. Further
examples of stable interactions can be found in \cite{KL15},
\cite{KL14}, \cite{KL16} for various higher-derivative models. In
all these models, the canonical energy is unbounded at free level,
while the stability is due to another bounded conserved quantity.

The Hamiltonian formalism for non-singular higher-derivative
theories was introduced by Ostrogradski \cite{Ostr}. Its
generalization for singular Lagrangians was first worked out in the
paper \cite{GLT}. The general constrained Hamiltonian formalism of
higher-derivative systems was further developed since that in
various directions. In particular it was adapted for
higher-derivative gravity in a series of works starting from
\cite{BL}, for recent developments and further references see
\cite{KOT}, \cite{OE}. Notice that all these reformulations are
connected by canonical transformations, so they cannot replace the
unbounded Hamiltonian with any bounded quantity. The canonical
Hamiltonian, being a canonical energy expressed in terms of phase
space variables, is always unbounded for non-degenerate
higher-derivative systems. Once the higher-derivative Lagrangian is
degenerate, the phase space variables are subject to constraints. On
the constraint surface, the canonical Hamiltonian can be bounded if
the constraints are strong enough. The examples of this phenomenon
are the same as previously mentioned cases of on-shell bounded
canonical energy. The paper \cite{KLS-m} demonstrates that once
Lagrange anchor is admitted by the equations of motion, the
first-order formalism of the theory admits a constrained Hamiltonian
formulation. If the model admits multiple Lagrange anchors, the
first-order formalism will be multi-Hamiltonian. Furthermore, it is
the conserved quantity connected to the time-shift symmetry by the
Lagrange anchor which serves as Hamilton function. With this regard,
the higher-derivative field theories of this class are expected to
admit multi-Hamiltonian formalism where some of Hamiltonians are
bounded. Once the classical Hamiltonian is bounded, the theory,
being canonically quantized with respect to corresponding Poisson
bracket, has a good chance to remain stable at quantum level.

The paper \cite{KLS14} provides a list of examples of
higher-derivative systems admitting  multiple Lagrange anchors,
including the PU oscillator. By the above mentioned reasons, every
model on this list has to be a multi-Hamiltonian system. It has been
earlier noticed that the free PU oscillator admits alternative
Hamiltonian formulations \cite{BK}, \cite{DS}. It has been observed
that the series of canonically inequivalent Hamiltonians includes
the bounded ones, while the canonical Ostrogradski Hamiltonian is
unbounded. Later, the multi-Hamiltonian formulations of PU
oscillator have been re-derived and re-interpreted from various
viewpoints in \cite{KL15}, \cite{BenM}, \cite{Ben}, \cite{Most},
\cite{Mas15}, \cite{Mas16}. All these observations can be summarized
in the statement that the PU oscillator of order $2n$ admits the
$n$-parameter series of alternative Hamiltonians and associated
Poisson brackets. Once the equations of motion admit Hamiltonian
formulation with bounded Hamilton functions, the dynamics is stable
classically and quantum-mechanically. It is also worse to notice
that the PU oscillator equation of motion admits the interaction
vertices such that do not spoil the classical stability
\cite{KLS14}, \cite{KLS16}. These vertices are non-Lagrangian, while
the interacting higher-derivative equations, being brought to the
first-order formalism, still remain Hamiltonian with positive
Hamilton function \cite{KL14}, \cite{KL15}. In this way, the PU
oscillator equation admits inclusion of interactions such that leave
the dynamics stable beyond the free level and admit Hamiltonian
formulation. Notice that the stability of PU oscillator with the
Lagrangian interaction vertices is studied once and again for
decades. In some cases, the model admits isles of stability, see
e.g. \cite{Smilga1}, \cite{Smilga2}, \cite{Pavsic1}, \cite{Pavsic2},
and \cite{PUPert} for the most recent results and review, while it
is unstable in general, unlike the case of above mentioned
non-Lagrangian interactions.

If the equations of motion admit a Lagrange anchor, the dynamics
have to admit a constrained Hamiltonian formulation \cite{KLS-m}.
With multiple Lagrange anchors, the dynamics should be
multi-Hamiltonian. In general, the construction of Hamiltonian
formulation for a given Lagrange anchor is implicit \cite{KLS-m}. A
direct relation between the Lagrange anchor and corresponding
Hamiltonian formalism has been established for the PU oscillator in
\cite{KL15}. In \cite{KLS14}, \cite{KLS16}, the interactions are
introduced, being compatible with the Lagrange anchor. The stable
interactions are found by means of the factorization method
\cite{KLS14} and proper deformation method \cite{KLS16}. These two
methods are equivalent \cite{KL16} in principle, though they apply
different techniques. Recently, more examples has become known of
stable interaction vertices in various higher-derivative models with
unbounded canonical energy at free level. The examples include PU
theory \cite{KLS14}, \cite{KL15}, \cite{Mas15}, \cite{Mas16},
Podolsky electrodynamics \cite{KLS14}, and higher-derivative
extensions of the Chern-Simons theory \cite{KKL}. The stable
interaction vertices are explicitly covariant in all the field
theoretical examples, though they do not follow from the least
action principle. The existence of a Lagrange anchor, however,
implies that these models have to admit the Hamiltonian description
at interacting level.

To the best of our knowledge, no explicit example has been known yet
of the higher-derivative field theory admitting multi-Hamiltonian
formulation. In this work, we construct the multi-Hamiltonian
formulation for higher-derivative extensions of Chern-Simons theory.
The canonical unbounded Hamiltonian is included into the
two-parametric series of admissible Hamilton functions. The series
can also include bounded Hamiltonians in some cases. The existence
of a bounded Hamiltonian depends on the parameters in the
third-order equations. We also demonstrate that the covariant
interactions exist such that the higher-derivative theory still
admits bounded Hamiltonian, and therefore it remains stable at
interacting level if the free model was stable.

We consider the class of theories of the vector field $A=A_\mu
dx^\mu$ in $3d$ Minkowski space with the free action functional
\begin{equation}\label{S-MCS}
S[A]=\frac{1}{2}\int \ast A\wedge ( m^2 \alpha_0 A+ m \alpha_1\ast
dA+a_2\ast d \ast d A+\alpha_3m^{-1}\ast d\ast d\ast d A+\ldots)\,.
\end{equation}
Here, $d$ is the de-Rham differential, $\ast$ is the Hodge star
operator, $m$ is a dimensional constant,
$\alpha_0,\alpha_1,\alpha_2,\alpha_3,\ldots$ are the dimensionless
constant real parameters. Depending on the values of the parameters,
the action (\ref{S-MCS}) can reproduce various $3d$ field theories,
including the Chern-Simons-Proca theory \cite{Townsend},
\cite{DJ1984}, topologically massive gauge theory \cite{tmgt1},
\cite{tmgt2},   Maxwell-Chern-Simons-Proca model \cite{MCS1},
\cite{MCS2}, Lee-Wick electrodynamics \cite{LW1}, \cite{LW2} and
extended Chern-Simons \cite{Deser}. The classical stability of the
model (\ref{S-MCS}) is considered in the works \cite{KKL},
\cite{KKL16}. It has been found that the model
 admits multiple conserved tensors being connected with the time translation by the Lagrange
 anchors. The anchors are the polynomials in the Chern-Simons operator $\ast d$.
 The set of conserved quantities can include bounded ones. This  depends on the roots of the characteristic equation
\begin{equation}\label{Char-MCS}
\alpha_0+\alpha_1z+\alpha_2
z^2+\alpha_3z^3+\ldots=0\,,
\end{equation}
 Here, $z$
is considered as a formal complex-valued variable, and $\alpha_k$
are the parameters of the model (\ref{S-MCS}). As is established in
\cite{KKL16}, the model (\ref{S-MCS}) admits a bounded conserved
tensor and, hence, it is stable iff all the non-zero simple roots of
equation (\ref{Char-MCS}) are real, while zero root may have the
maximal multiplicity 2, and no roots occur with a higher
multiplicity.

In this paper, we focus at the model (\ref{S-MCS}) with at maximum
third-order derivatives, i.e. the action reads
\begin{equation}\label{S-MCS3}
S[A]=\frac{1}{2}\int \ast A\wedge (  \alpha_1 m\ast dA+\alpha_2\ast
d \ast d A+m^{-1}\ast d\ast d\ast d A)\,,
\end{equation}
with $\alpha_1,\alpha_2$ being two independent dimensionless
parameters. This model has been proposed in \cite{Deser} as the
third-order extension of Chern-Simons theory.  The model is
obviously gauge invariant. We construct the multi-Hamiltonian
constrained formalism for this model at free level.  For similar
reasons, the more general case (\ref{S-MCS}) has to be a
multi-Hamiltonian system, with a broader class of admissible
Hamiltonians depending on the structure of roots in
(\ref{Char-MCS}). As the construction of multi-Hamiltonian formalism
becomes more cumbersome with growth of the order of derivatives, we
do not go beyond the third-order models in this paper. \footnote{The
multi-Hamiltonian formulation for the gauge-invariant extension of
the Chern-Simons model of the fourth order have been constructed in
\cite{AKL17r}.}

Let us explain what do we understand by constrained
multi-Hamiltonian formalism.  At first, notice the obvious fact that
the higher-derivative field equations can be always reduced to the
first-order derivatives in time by introducing extra fields
absorbing higher the time derivatives. The first-order equations are
said multi-Hamiltonian if there exists $k$-parametric series of
Hamiltonians $H(\beta, \varphi,\nabla\varphi, \nabla^2\varphi,
    \nabla^3\varphi,  \ldots)$ and Poisson brackets $\{\varphi^a(\vec{x}),\varphi^b(\vec{y})\}_\beta$,
with $\beta_1, \ldots , \beta_k$ being constant parameters and
$\nabla$ denoting derivatives by space $\vec{x}$, such that the
equations constitute constrained Hamiltonian system with any
$\beta$, i.e.
\begin{equation}\label{multi-H-gen}
 \dot{\varphi}{}^a= \{ \varphi^a, H_{T(\beta)} \}_\beta \, ,
 \end{equation}
 $$
 H_{T}(\beta)= H(\beta, \varphi,\nabla\varphi, \nabla^2\varphi,
    \nabla^3\varphi,  \ldots) +\lambda^AT_A(\varphi,\nabla\varphi, \nabla^2\varphi,
    \nabla^3\varphi,  \ldots) \, ;
    $$
\begin{equation}\label{T-gen}
 T_A(\varphi,\nabla\varphi, \nabla^2\varphi,
    \nabla^3\varphi,  \ldots) =0\, .
    \end{equation}
The rhs of equations (\ref{multi-H-gen}) does not depend on the
parameters $\beta$, while both the total Hamiltonian $H_{T}(\beta)$
and the Poisson bracket do. In the other wording, changing values of
parameters $\beta$, we simultaneously change Hamiltonian
$H_{T}(\beta)$ and Poisson brackets $\{\cdot , \cdot \}_\beta$ in
such a way that the equations of motion (\ref{multi-H-gen}) remain
intact.

Any higher-derivative Lagrangian field theory always admits at least
one Hamiltonian formulation which can be constructed by the
Ostrogradski method in the unconstrained case, and by various
generalizations \cite{GLT}, \cite{BL}, \cite{KOT}, \cite{OE}
developed for the constrained systems. In this paper, we develop the
Hamiltonian formalism of higher-derivative field theory in several
respects by the example of the model (\ref{S-MCS3}). At first, the
third-order extension of the Chern-Simons model (\ref{S-MCS3}) is
shown to admit a two-parameter series of constrained Hamiltonian
formulations. The Hamiltonians from this series can be bounded from
below in some cases, depending upon parameters $\alpha_1,\alpha_2$,
even though Ostrogradski's Hamiltonian of the model is unbounded in
all the instances. The second is that the free higher-derivative
equations of this model admit inclusion of covariant interactions
which do not break the stability if the theory have bounded
conserved quantity at free level. Furthermore, the stable theory
admits constrained Hamiltonian formulation at interacting level with
a bounded Hamilton function.

Let us also remark that the multi-Hamiltonian formulation helps to
resolve the discrepancy between classical stability of
higher-derivative dynamics and quantum instability which is
connected to the unboundedness of canonical Hamiltonian. As it is
noticed in \cite{KKL}, the stable higher-derivative extensions of
the Chern-Simons model realize the reducible representations which
are decomposed into the unitary irreps in some cases. In the other
cases, the representations are non-unitary or non-decomposable. If
the model admits only unbounded conserved tensors, it corresponds to
a non-unitary representation, while the models with unitary
representations admit non-Ostrogradski's bounded Hamiltonians. If
the theory is quantized with the bounded Hamiltonian, and the
commutation relations are imposed in accordance with the
corresponding Poisson brackets, the theory will be
quantum-mechanically stable, as it is at the classical level.

Let us make some comments on the interactions which do not break the
stability of the higher-derivative theory. An example of stable
couplings in the model (\ref{S-MCS}) has been noticed in \cite{KKL}
in the case involving massive Proca term, so it is the theory
without gauge symmetry. In the present paper, we consider the gauge
model (\ref{S-MCS3}) and introduce gauge-invariant interaction with
spinors. This class of interactions can be viewed as a
generalization to the non-minimal stable couplings of $d=4$ Podolsky
electrodynamics to the spinor matter proposed in ref. \cite{KLS14}.

The article is organized as follows. In Section 2, we describe
conserved tensors of the third-order model (\ref{S-MCS3}). We also
relate the existence of bounded conserved tensors with the structure
of the corresponding Poincar\'e group representation. In doing that,
we mostly follow the general prescriptions of \cite{KKL} and
\cite{KKL16}. The section is self-contained, however. In Section 3,
the multi-Hamiltonian formulation is constructed with the
Hamiltonians defined by the conserved tensors of the Section 2. In
Section 4, we introduce the interactions with spin $1/2$ such that
do not break the stability of higher-derivative theory if the theory
is stable at free level. After that, we demonstrate that the
higher-derivative interacting theory still admits Hamiltonian
formulation in all the instances, even if the vertices are not
Lagrangian.

\section{Conserved tensors}
\noindent For the action (\ref{S-MCS3}), the Lagrange equations read
\begin{equation}\label{EL-MCS3}
\frac{\delta S}{\delta A}\equiv(\alpha_1 m \ast d+\alpha_2\ast d
\ast d+\frac{1}{m}\ast d \ast d\ast d)A=0\,.
\end{equation}
The third-order time derivatives are involved in these equations.
That is why, the conserved quantities can involve the second-order
time derivatives.

The equations (\ref{EL-MCS3}) correspond to reducible representation
of the Poincar\'e group. Specifics of the representation depend on
the constants $\alpha_1,\alpha_2$. Different cases are distinguished
by the structure of roots in the characteristic equation
\begin{equation}\label{CE}
z^3+\alpha_2z^2+\alpha_1z=0\,
\end{equation}
associated to the field equations (\ref{EL-MCS3}). Here, $z$ is a
formal unknown variable, and $\alpha_1, \alpha_2$ are the parameters
of the model.
There are the following different cases distinguished by the structure of roots for the variable $z$: \\
\begin{equation}\label{Cases}\begin{array}{lll}
\text{(A)} &\quad\alpha_1\neq0\,,\qquad \alpha_2{}^2-4\alpha_1>0, &\text{two simple real nonzero roots, and one simple zero root;} \\
\text{(B)} &\quad\alpha_1=0\,,\qquad \alpha_2\neq0, & \text{one simple real nonzero root, and one zero root of multiplicity two};\\
\text{(C)} &\quad\alpha_1\neq0\,,\qquad \alpha_2{}^2-4\alpha_1=0,& \text{one real nonzero root of multiplicity two, and one simple zero root};\\
\text{(D)} &\quad\alpha_1=0\,,\qquad \alpha_2=0, & \text{one zero root of multiplicity three};\\
\text{(E)} &\quad\alpha_1\neq0\,,\qquad \alpha_2{}^2-4\alpha_1<0, &\text{two simple complex conjugate roots, and one simple zero root}.
\end{array}\end{equation} In cases A and B, the representation is unitary
and reducible. In case A, the representation is decomposed into two
irreducible sub-representations. Each one corresponds to a self-dual
massive spin 1, while the masses can be different. In case B, the
set of sub-representations includes a massless spin 1 and a massive
spin 1 subject to a self-duality condition. Cases C and D correspond
to reducible indecomposable non-unitary representations. These two
options are distinguished by different multiplicity of the multiple
real root in eq. (\ref{CE}). In case E, the representation is
irreducible and non-unitary. So, one can see that the field
equations (\ref{EL-MCS3}) can describe either unitary or non-unitary
representations of the $3d$ Poincar\'e group depending on the
relations between the parameters $\alpha_1,\alpha_2$.

The third-order field equations (\ref{EL-MCS3}) admit two-parameter
series of on-shell conserved second-rank tensors
\begin{equation}\label{Tmunu}
T_{\mu\nu}(\beta_1,\beta_2)=\beta_1(T_1)_{\mu\nu}+\beta_2(T_2)_{\mu\nu}\,,
\end{equation}
where $\beta_1,\beta_2$ are the real constant parameters, and
$(T_a)_{\mu\nu},a=1,2$ read
\begin{equation}\label{T12munu}\begin{array}{l} \displaystyle
(T_{1})_{\mu\nu}=\frac{1}{m}\Big\{(G_\mu F_\nu+G_\nu
F_\mu-\eta_{\mu\nu}G_\rho F^\rho)+\alpha_2
m(F_{\mu}F_{\nu}-\frac12\eta_{\mu\nu}F_{\rho}F^{\rho})\Big\}\,,\\[3mm]\displaystyle
(T_{2})_{\mu\nu}=\frac{1}{m^2}\Big\{(G_\mu
G_\nu-\frac12\eta_{\mu\nu} G_\rho
G^\rho)-\alpha_1m^2(F_{\mu}F_{\nu}-\frac12\eta_{\mu\nu}F_{\rho}F^{\rho})\Big\}\,.
\end{array}\end{equation}
Here we use the notation\footnote{The Minkowski metric is taken with
mostly negative signature.}
\begin{equation}\label{FGinv}
F_\mu\equiv\varepsilon_{\mu\rho\nu}\partial^\rho A^\nu=(\ast
dA)_\mu\,,\qquad G_\mu\equiv\partial_\mu\partial^\nu A_\nu-\Box
A_\mu=(\ast d\ast dA)_\mu\,,\qquad \varepsilon_{012}=1\,.
\end{equation}
Tensor $T_1$ is a canonical energy-momentum for the action
(\ref{S-MCS3}), while $T_2$ is another independent conserved
quantity. As $F$ and $G$ are gauge invariant quantities, the tensor
(\ref{Tmunu}) is gauge invariant with any $\beta$. Also notice that
$F_i,G_i, i=1,2$ define independent unconstrained Cauchy data for
the field equations (\ref{EL-MCS3}). Once $T_1$ is linear in $G$, it
is unbounded anyway. The general entry of the series (\ref{Tmunu})
is bilinear in both $G$ and $F$. So, $T(\beta)$ can be bounded, in
principle, if $\beta_2\neq 0$.

The conserved tensors of the series (\ref{Tmunu}) are connected to
the invariance of the model with respect to the space-time
translations if the parameters meet the condition
\begin{equation}\label{Delta}
\beta_1^2-\alpha_2\beta_1\beta_2+\alpha_1\beta_2^2\neq0\,.
\end{equation}
This connection can be traced by the Lagrange anchor method along
the same lines as in the paper \cite{KKL16}. From this perspective,
any representative of the series (\ref{Tmunu}) satisfying condition
(\ref{Delta}) can be viewed as energy-momentum.

The $00$-component of the conserved tensor $T(\beta_1,\beta_2)$ from
the series (\ref{Tmunu}) can be bounded or unbounded from below
depending of the parameters $\alpha$ involved in the equations
(\ref{EL-MCS3}) and on specific values of $\beta$. Once the
representation is unitary (that corresponds to the cases A,B in
classification (\ref{Cases})), the bounded representatives exist
with certain $\beta$'s, as we shall see in the next section. For
non-unitary representations (the cases C,D,E), the $00$-component of
the conserved tensor $T(\beta)$ is unbounded in all the instances.
As the existence of bounded conservation law provides the classical
stability of the model, the theory is stable if the parameters of
the model meet the conditions (\ref{Cases}.A) or (\ref{Cases}.B),
and it is unstable in all the other cases. Also notice that the
canonical energy $(T_1)_{00}$ is always unbounded.

The conserved tensors are defined modulo on-shell vanishing terms.
So, we have equivalence classes of conserved quantities which
coincide on-shell, being off-shell different. The choice of specific
representative of the equivalence class is a natural ambiguity in
the definition of conserved quantity. We mention this ambiguity
 because it has a natural counterpart in the Hamiltonian formalism considered in the next section.
 As far as the linear equations (\ref{S-MCS3}) admit bilinear gauge invariant
conserved tensors (\ref{Tmunu}), it is natural to consider the
series up to quadratic on-shell vanishing terms. The most general
gauge-invariant bilinear and symmetric representative in the
equivalence class of $T_{\mu\nu}(\beta_1,\beta_2)$ (\ref{Tmunu})
reads
\begin{equation}\label{Tmunurd}
T_{\mu\nu}(\beta_1,\beta_2,\beta_3,\beta_4)=T_{\mu\nu}(\beta_1,\beta_2)+\frac{\beta_3}{2m}
\Big(F_{\mu}\frac{\delta S}{\delta A^{\nu}}+F_{\nu}\frac{\delta S}{\delta A^{\mu}}\Big)+
\frac{\beta_4}{2m^2}\Big(G_{\mu}\frac{\delta S}{\delta A^{\nu}}+G_{\nu}\frac{\delta S}{\delta A^{\mu}}\Big)\,,
\end{equation}
Two real parameters $\beta_3,\beta_4$ label different
representatives of the same equivalence class of conserved tensors,
while $\beta_1,\beta_2$ determine the equivalence class of conserved
tensor as such. Only one of two constants $\beta_3,\beta_4$ is
independent. The other one can be absorbed by the multiplication of
the equations of motion by the constant overall factor.

In the next section, we construct a multi-Hamiltonian formulation
where $00$-components of the conserved tensors $T_{\mu\nu}(\beta_1,\beta_2,\beta_3,\beta_4)$
(\ref{Tmunurd}) serve as Hamiltonians, and all the  values of the
parameters $\beta_1$ and $\beta_2$, being subject to condition (\ref{Delta}), are admissible. We consider all
the cases in a uniform way, be the Hamiltonian bounded or not.

\section{Multi-Hamiltonian formulation}
The multi-Hamiltonian formalism is constructed for the equations
(\ref{EL-MCS3}) in three steps. First, the higher-derivative
equations are reduced to the first-order in time by introducing
extra variables to absorb the time derivatives of the original field
$A$. The first-order equations are split in two subsets. The first
one includes the evolutionary type equations, while the other
equations are the constraints. The latter ones do not involve the
time derivatives of the fields. Second, the zero-zero component of
the most general conserved tensor of the series (\ref{Tmunu}) is
taken as the Hamiltonian of the model. As far as the considered
model is constrained, the Hamiltonian involves a linear combination
of constraints. Third, the series of Poisson bracket is found for
the series of Hamiltonians such that the evolutionary-type equations
of motion take the constrained multi-Hamiltonian form
(\ref{multi-H-gen}).

Let us reduce the third-order field equations  (\ref{EL-MCS3}) to
the first order in time $x^0$. Introduce new fields absorbing the
first and second order time derivatives of original field $A_i,
i=1,2$, while the time derivatives of $A_0$ eventually drop out from
the equations. We chose the gauge-invariant quantities $F_i$, $G_i$,
$i=1,2$ (\ref{FGinv}) as new variables  absorbing the time
derivatives of $A$,
\begin{equation}\label{varphi}
F_i=\varepsilon_{ij}(\dot{A}_j-\partial_jA_0)\,,\qquad
G_i=-\ddot{A}_i+\partial_i\dot{A}_0+\partial_{j}(\partial_jA_i-\partial_iA_j)\,,\qquad
i,j=1,2\,,
\end{equation}
with $\varepsilon_{ij}=\varepsilon_{0ij}$ being the $2d$ Levi-Civita
symbol. Substituting these variables into (\ref{EL-MCS3}), we arrive
at the following first-order equations in terms of the fields
$A_\mu, F_i, G_i$ :
\begin{equation}\label{dvarphi}\begin{array}{l}
\dot{A}_i=\partial_iA_0-\varepsilon_{ij}F_j\,,\\[3mm]
\dot{F}_i=\varepsilon_{ij}\bigr[\partial_k(\partial_kA_j-\partial_jA_k)-G_j\bigl]\,,\\[3mm]
\dot{G}_i=\varepsilon_{ij}\bigr[\partial_k(\partial_kF_j-\partial_jF_k)+m(\alpha_2G_j+\alpha_1mF_j)\bigl]\,,\\[3mm]
\end{array}\end{equation}
\begin{equation}\label{Theta}
\displaystyle\Theta\equiv\varepsilon_{ij}\partial_i\biggr(\frac{1}{m}G_j+\alpha_2F_j+\alpha_1mA_j\biggl)=0\,.
\end{equation}
In terms of fields $A,F,G$, the evolutionary equations
(\ref{dvarphi}) represent the first-order form of the space
components of the Lagrange equations (\ref{EL-MCS3}). The zero
component of the original field equations (\ref{EL-MCS3}) is a
constraint (\ref{Theta}), which does not involve time derivatives.
Once the constraint $\Theta$ conserves with account for the
evolutionary equations, no secondary constraints are imposed on the
fields. The first-order equations (\ref{dvarphi}), (\ref{Theta}) are
obviously equivalent to the original third-order ones
(\ref{EL-MCS3}).

In the first-order formalism, the equations are invariant under the
gauge transformation
\begin{equation}\label{gt}
\delta_\xi A_0=\partial_0\xi(x)\,,\qquad\delta_\xi
A_i=\partial_i\xi(x)\,,\qquad \delta_\xi F_i=\delta_\xi G_i=0\, ,
\end{equation}
where $\xi$ is the gauge transformation parameter, being arbitrary
function of $x$. In what follows, it is natural to consider the
field $A_0$ as the Lagrange multiplier associated to the constraint
(\ref{Theta}). This interpretation is consistent with the gauge
transformation (\ref{gt}) which includes the time derivative of the
gauge parameter, as it should be for Lagrange multiplier in the
constrained Hamiltonian formalism.

In the first-order formalism, the zero-zero component of the
conserved tensor (\ref{Tmunurd}) reads
\begin{equation}\label{H0}\begin{array}{c}\displaystyle
T_{00}(\beta_1,\beta_2)=\frac{1}{2m^2} \biggr\{\beta_2
\bigr[\partial_iF_j(\partial_iF_j-\partial_jF_i)+(G_i)^2\bigl]+
2m\beta_1\bigr[\partial_iF_j(\partial_iA_j-
\partial_jA_i)+G_iF_i)\bigl]+\\[3mm]\displaystyle+
m^2(\beta_1\alpha_2-\beta_2\alpha_1)
\bigr[\partial_iA_j(\partial_iA_j-\partial_jA_i)+(F_i)^2\bigl]\biggr\}\,.
\end{array}\end{equation}
We treat this quantity as the series of on-shell Hamiltonians
parameterized by constants $\alpha,\beta$. Off-shell, the
Hamiltonian can be a sum of (\ref{H0}) and constraints. We chose the
following ansatz for the total Hamiltonian:
\begin{equation}\label{H}\begin{array}{l}\displaystyle
H_T(\beta_1,\beta_2,\gamma)\equiv T_{00}(\beta_1,\beta_2)+\\[3mm]\displaystyle\qquad+
\Big[\frac{\beta_1^2-\alpha_2\beta_1\beta_2+\alpha_1\beta_2^2}{\beta_1-\alpha_2\beta_2-\alpha_1\gamma}A_0+
\frac1m\frac{\beta_1\beta_2+\alpha_1\beta_2\gamma-\alpha_2\beta_1\gamma}{\beta_1-\alpha_2\beta_2-\alpha_1\gamma}\varepsilon_{ij}\partial_iA_j
+\frac{1}{m^2}\frac{\beta_2^2+\beta_1\gamma}{\beta_1-\alpha_2\beta_2-\alpha_1\gamma}\varepsilon_{ij}\partial_iF_j\Big]\Theta\,,
\end{array}\end{equation}
where $\beta_1\,,\beta_2\,,\gamma$ are constant parameters. On
account of the constraint (\ref{Theta}), the quantities (\ref{H0})
and (\ref{H}) coincide on shell. The parameter $\gamma$  is
introduced to control the inclusion of the constraint term into the
Hamiltonian\footnote{The constraint terms, being included into
Hamiltonian, do not contribute to the equations of motion for the
gauge-invariant quantities. These terms, however, can alter the
equations for the non-gauge-invariant quantities. We are seeking for
a series of the Hamiltonians and Poisson brackets such that
literally reproduce the first-order form (\ref{dvarphi}) of the
original third-order equations (\ref{S-MCS3}) for all the variables,
including the original vector field. As $A_\mu$ is not a gauge
invariant, we keep the constraint terms under control in the
Hamiltonian.}. The admissible values of the parameters $\beta$ and
$\gamma$ subject to conditions
\begin{equation}\label{gammacond}
\beta_1^2-\alpha_2\beta_1\beta_2+\alpha_1\beta_2^2\neq0\,,\qquad \beta_1-\alpha_2\beta_2-\alpha_1\gamma\neq0\,.
\end{equation}
Here, the first condition implies that the conserved quantity
(\ref{H0}) is connected to the invariance of the model
(\ref{EL-MCS3}) with respect to the time translations, see eq.
(\ref{Delta}). Both the conditions (\ref{gammacond}) ensure that the
numerical factor at the Lagrange multiplier $A_0$ in the Hamiltonian
(\ref{H}) is nonzero and nonsingular. Once these two requirements
are met,  any conserved quantity (\ref{H0}) can serve as the
Hamiltonian with appropriate Poisson bracket.

Now, let us seek for the Poisson brackets among the fields $A_i,
F_i,G_i, i=1,2$ such that the equations (\ref{dvarphi}),
(\ref{Theta}) take the constrained multi-Hamiltonian form
(\ref{multi-H-gen}), (\ref{T-gen}) with the Hamiltonian defined by
relations (\ref{H0}), (\ref{H}) and the constraint (\ref{Theta}).
Given the series of Hamiltonians  (\ref{H0}), (\ref{H}) and the
r.h.s. of the equations (\ref{dvarphi}), we arrive at the system of
linear algebraic equations defining the series of Poisson brackets
$\{\cdot , \cdot \}_{\beta,\gamma}$ \, :
\begin{equation}\label{PB}\begin{array}{l}\displaystyle
\{A_i, H_T(\beta,\gamma)
\}_{\beta,\gamma}=\partial_iA_0-\varepsilon_{ij}F_j\,,\\[3mm]
\{F_i,H_T(\beta,\gamma)\}_{\beta,\gamma}=\varepsilon_{ij}\bigr[\partial_k(\partial_kA_j-\partial_jA_k)-G_j\bigl]\,,\\[3mm]
\{G_i,
H_T(\beta,\gamma)\}_{\beta,\gamma}=\varepsilon_{ij}\bigr[\partial_k(\partial_kF_j-\partial_jF_k)+m(\alpha_2G_j+\alpha_1mF_j)\bigl]\,.
\end{array}\end{equation}
The Poisson bracket, being defined by these equations, involves five
 independent parameters $\alpha_1, \alpha_2, \beta_1,\beta_2,\gamma$. The bracket
eventually reads
\begin{equation}\label{PBsol}\begin{array}{l}\displaystyle
\{G_i(\vec{x}),G_j(\vec{y})\}_{\beta,\gamma}=m^3
\frac{(\alpha_1-\alpha_2^2)\beta_1+\alpha_1\alpha_2\beta_2}
{\beta_1^2-\alpha_2\beta_1\beta_2+\alpha_1\beta_2^2}
\varepsilon_{ij}\delta(\vec{x}-\vec{y})\,,
\\[3mm]\displaystyle
\{F_i(\vec{x}),G_j(\vec{y})\}_{\beta,\gamma}=m^2\frac{\alpha_2\beta_1-\alpha_1\beta_2}
{\beta_1^2-\alpha_2\beta_1\beta_2+\alpha_1\beta_2^2}
\varepsilon_{ij}\delta(\vec{x}-\vec{y})\,,\\[3mm]\displaystyle
\{F_i(\vec{x}),F_j(\vec{y})\}_{\beta,\gamma}=\{A_i(\vec{x}),G_j(\vec{y})\}_{\beta,\gamma}=
\frac{-m\beta_1}{\beta_1^2-\alpha_2\beta_1\beta_2+\alpha_1\beta_2^2}\varepsilon_{ij}\delta(\vec{x}-\vec{y})\,,\\[3mm]\displaystyle
\{A_i(\vec{x}),F_j(\vec{y})\}_{\beta, \gamma}=\frac{\beta_2}{\beta_1^2-\alpha_2\beta_1\beta_2+\alpha_1\beta_2^2}\varepsilon_{ij}\delta(\vec{x}-\vec{y})\,,\\[3mm]\displaystyle
\{A_i(\vec{x}),A_j(\vec{y})\}_{\beta,\gamma}=\frac{1}{m}\frac{\gamma}{\beta_1^2-\alpha_2\beta_1\beta_2+\alpha_1\beta_2^2}\varepsilon_{ij}\delta(\vec{x}-\vec{y})\,.
\end{array}\end{equation}
 The accessory parameter $\gamma$ controls the constraint terms in the
total Hamiltonian (\ref{H}). As is seen, the same parameter defines
the Poisson bracket between the components $A_i$ of gauge potential.
This parameter does not contribute to the Poisson brackets between
the physical observables, being the functions of the gauge-invariant
quantities $F_i,G_i$, and the strength
$\varepsilon_{ij}\partial_iA_j$. That is why, $\gamma$ can be
considered as an accessory parameter. Inclusion of $\gamma$-terms
into total Hamiltonian and brackets allows us to literally reproduce
in Hamiltonian form the first-order dynamical equations
(\ref{dvarphi}) for all the quantities, be they gauge-invariant or
not.

Let us make one more comment on the meaning of the accessory
parameter $\gamma$ which defines the bracket between $A_i$ and does
not affect on the brackets of gauge-invariant quantities. Notice
that the Poisson brackets in gauge theory have the inherent
ambiguities. The general study of these ambiguities can be found in
ref. \cite{LS05}. In context of the bracket (\ref{PBsol}), one of
these ambiguities turns out relevant. It is related to the option of
redefining the Poisson bracket by adding the bi-vector, being the
wedge product of gauge symmetry generator operator to another
vector. This redefinition does not affect the brackets between
gauge-invariant observables, while it can alter the brackets of
non-gauge-invariant quantities. The bracket (\ref{PBsol}) involves
the ambiguous terms of this type, and it is the ambiguity which is
controlled by the accessory parameter $\gamma$.

The problem of identification of ambiguous terms in the Poisson
bracket is a subtle issue. The Poisson bracket (\ref{PBsol}) is
ultralocal between components of $A_i$ with no derivatives involved,
while the generator of the gauge symmetry for $A_i$ (\ref{gt})
involves a derivative. Thus, the ambiguous terms in the Poisson
bracket cannot be absorbed by adding the wedge product of the gauge
symmetry generator, being a derivative,  to another vector, being a
polynomial in the partial derivatives $\partial_i$. The problem is
solved by including the inverse Laplace operator
$\Delta^{-1}=(\partial_i\partial_i)^{-1}$ into the coefficient at
the gauge generator. The space non-locality of this type is usually
considered as admissible for the constrained Hamiltonian formalism
in the field theory\footnote{For example, the inverse Laplacian
contributes to the Dirac brackets of vector potential  to electric
strength in the Maxwell electrodynamics in the Coulomb gauge.  }. To
represent the bracket (\ref{PBsol}) between the components of $A_i$
in terms of gauge generators, we use the following identical
representation for the $2d$ Levi-Civita tensor $\varepsilon_{ij}$:
\begin{equation}\label{2dLeviCivita}
\varepsilon_{ij}=\frac{1}{2\Delta}(\varepsilon_{im}\partial_m\partial_j-\varepsilon_{jm}\partial_m\partial_i)\,.
\end{equation}
Substituting $\varepsilon_{ij}$ from this relation into rhs of the Poisson bracket for the potential components, we rewrite the bracket in the form
\begin{equation}\label{r}
\{A_i(\vec{x}),A_j(\vec{y})\}_{\beta,\gamma}=\frac{1}{2}(V_i(\gamma)\partial_j-V_j(\gamma)\partial_i)\delta(\vec{x}-\vec{y})\,,\qquad V_i(\gamma)=\frac{\gamma}{m(\beta_1^2-\alpha_2\beta_1\beta_2+\alpha_1\beta_2^2)}\frac{\varepsilon_{im}\partial_m}{\Delta}.
\end{equation}
Here, all the partial derivatives act on argument $\vec{x}$ in the
delta-function. Once the operator $\partial_i$ is a gauge generator
for the field $A_i$, the vector $V_j(\gamma)$ parametrizes the
ambiguity in the Poisson bracket. Thus, we treat the parameter
$\gamma$ as inherent ambiguity of Poisson bracket in the gauge
theory outlined in ref. \cite{LS05}.

Let us summarize all the aspects related to the ambiguity in
parametrization of the multi-Hamiltonian formulation of the
equations (\ref{EL-MCS3}). The Hamiltonian and brackets (\ref{H}),
(\ref{PBsol}) involve $5$ parameters. Two of them, $\alpha_1$ and
$\alpha_2$, define the original equations (\ref{EL-MCS3}). The
constants $\beta_1, \beta_2$ parameterize the series of conserved
tensors tensors (\ref{Tmunu}). These tensors admit gauge-invariant
re-definitions by on-shell vanishing terms (\ref{Tmunurd}), with one
more parameter in control of corresponding ambiguity. The zero-zero
components of the conserved tensors are chosen as  Hamiltonians for
the first-order formulation (\ref{dvarphi}), (\ref{Theta}) of the
original third-order equations (\ref{EL-MCS3}). In this way, the
ambiguity in the off-shell definition of the conserved tensors is
converted into the ambiguity in the constraint terms of the
Hamiltonian. The later ambiguity does not contribute to the
equations of motion for the gauge-invariant quantities $F_i,G_i,
\varepsilon_{ij}\partial_iA_j$, while the equations of motion for
the potential components $A_i$ can alter. We seek for a series of
the Hamiltonians and Poisson brackets such that the Hamiltonian
equations literally reproduce the first-order form (\ref{dvarphi})
of the original third-order equations (\ref{S-MCS3}) for all the
variables, including the original vector field. In this case, one
and the same parameter has to control the ambiguity in the
Hamiltonian and Poisson bracket. It is the parameter $\gamma$. In
the free theory, $\gamma$ can be set to an arbitrary value. This
corresponds to the choice of the representative in the equivalence
class in the series of Hamiltonian formulations with the Hamiltonian
(\ref{H0}), (\ref{H}) and Poisson bracket (\ref{PBsol}). Thus,
$\gamma$ is an accessory parameter in the series of Hamiltonian
formulations unless the interaction is introduced. We keep $\gamma$
in the Hamiltonian formulation throughout this section to have the
contact with Section 4,  where the couplings  are introduced with
spinors. As we will see, this parameter becomes essential for
inclusion of consistent interactions in the non-linear model.

The Hamiltonians in the series (\ref{H}) can be bounded or unbounded
form below. The Hamiltonian  $H_T(\beta,\gamma)$ is on-shell bounded
if the parameters meet the conditions
\begin{equation}\label{b1b2}
\beta_1>0\,,\qquad
\beta_1^2-\alpha_2\beta_1\beta_2+\alpha_1\beta_2^2<0\,.
\end{equation}
In cases A,B in classification (\ref{Cases}), these conditions can
be satisfied by appropriate values of the parameters $\beta_1$,
$\beta_2$. In cases C, D, E of classification (\ref{Cases}),
conditions (\ref{b1b2}) are inconsistent. As we see, the bounded
Hamiltonians are included in the series (\ref{H}) once the equations
(\ref{EL-MCS3}) transform under unitary representations of the
Poincar\'e group. For non-unitary representations, every Hamiltonian
is unbounded in the series. We finally notice that condition
(\ref{b1b2}) is more restrictive than (\ref{Delta}). Thus, any
bounded conserved quantity serves as a Hamiltonian. The Ostrogradski
Hamiltonian, being included in the series (\ref{H}) with $\beta_1=1,
\beta_2=0$, is always unbounded.

For every $\beta,\gamma$, the Poisson bracket (\ref{PBsol}) is a
non-degenerate tensor, so it has an inverse, being a symplectic
two-form. The latter defines the series of Hamiltonian action
functionals
\begin{equation}\label{SHam}\begin{array}{c}\displaystyle
S(\beta,\gamma)=\int
\Big\{\frac{\beta_1^2-\alpha_2\beta_1\beta_2+\alpha_2\beta_2^2}{\beta_1-\alpha_2\beta_2-\alpha_1\gamma}(\alpha_1mA_i+2\alpha_2F_i+\frac{2}{m}G_i)
\varepsilon_{ij}\dot{A}_j+\frac{1}{m}\frac{\beta_1^2+((\alpha_2^2-\alpha_1)\beta_1-\alpha_1\alpha_2\beta_2)\gamma}{\beta_1-\alpha_2\beta_2-\alpha_1\gamma}
\varepsilon_{ij}F_i\dot{F}_j+
\\[3mm]\displaystyle+\frac{2}{m^2}\frac{\beta_1\beta_2+(\alpha_2\beta_1-\alpha_1\beta_2)\gamma}{\beta_1-\alpha_2\beta_2-\alpha_1\gamma}\varepsilon_{ij}G_i\dot{F}_j
+\frac{1}{m^3}\frac{\beta_2^2+\beta_1\gamma}{\beta_1-\alpha_2\beta_2-\alpha_1\gamma}\varepsilon_{ij}G_i\dot{G}_j
-H_{T}(\beta,\gamma)\Big\}d^3x\,,
\end{array}\end{equation}
where $H_T(\beta,\gamma)$ denotes the total Hamiltonian (\ref{H}).

For $\beta_1=1, \beta_2=\gamma=0$, we get Ostrogradski's action for the
variational model (\ref{S-MCS3}):
\begin{equation}\label{SHam0}\begin{array}{c}\displaystyle
S_{\text{Canonical}}=\displaystyle\int
\Big\{(\alpha_1mA_i+2\alpha_2F_i+\frac{2}{m}G_i)\varepsilon_{ij}\dot{A}_j-
\frac{1}{m}\varepsilon_{ij}F_i\dot{F}_j-A_0\Theta-(T_1)_{00}\Big\}d^3x\,,
\end{array}\end{equation}
where $(T_1)_{00}$ is the $00$-component of the canonical
energy-momentum tensor. The formula (\ref{SHam0}) follows from
(\ref{SHam}) for all the values of parameters $\alpha_1,\alpha_2$ of
the model (\ref{EL-MCS3}).

For $\beta_2\neq0$, we get the non-canonical Hamiltonian actions
that still result to the same original equations (\ref{EL-MCS3}).
Different actions in the series (\ref{SHam}) are not connected by a
canonical transformation. This is obvious because the Hamiltonian in
the series (\ref{H}) can be bounded from below, while the canonical
Hamiltonian (\ref{SHam0}) is always unbounded.

The Poincar\'e invariance can be questioned of the non-canonical
Hamiltonian actions (\ref{SHam}), and hence the covariance of the
corresponding quantum theory may seem in question. We do not
elaborate on this issue here, while we claim that the quantum theory
associated to any model in the series (\ref{SHam}) is
Poincar\'e-invariant. The argument is that the original
higher-derivative theory admits the series of covariant Lagrange
anchors \cite{KKL}. It is the series of anchors which underlies the
multi-Hamiltonian formulation (\ref{SHam}). One more reason is
provided by the fact that every Hamiltonian in the series (\ref{H0})
is $00$-component of the second rank tensor (\ref{Tmunurd}). All the
entries of the series transform in the same way, including
Ostrogradski's Hamiltonian.

\section{Stable interactions with spinor field}

As we have seen above, the higher-derivative extensions of the
Chern-Simons theory admit multi-Hamiltonian formulations. In some
cases, the Hamiltonians are bounded. In this section, we provide an
example of coupling to spinors such that the theory still has
bounded Hamiltonian and therefore it remains stable at interacting
level.

In \cite{KLS14}, the stable interaction is included for the
higher-derivative Podolsky's electrodynamics in the dimension $d=4$.
The stable interaction is non-Lagrangian in $d=4$, while the
Hamiltonian formalism is not considered there. So, the possibility
could be questioned of the canonical quantization of the interacting
model even without gauge invariance. The three-dimensional model
admits more options than its four-dimensional counterpart, because
(due to the presence of the Chern-Simons term) it can describe a
variety of reducible representations of the $3d$ Poincar\'e group.
Below we introduce the interaction mostly following the lines of
\cite{KL14} with regard to the $d=3$ specifics, and then we
construct the Hamiltonian formalism for the interacting theory.

Let us introduce coupling of the vector field $A$ and $2$-component
spinor field $\psi_a, a=1,2$ ($\overline{\psi}_a$ stands for
conjugate spinor) by imposing the following non-linear field
equations
\begin{equation}\label{int}\begin{array}{c}\displaystyle
\frac{\delta S\,\,}{\delta A^\mu}-J_\mu (\overline{\psi},\psi)\equiv
\varepsilon_{\mu\rho\nu}\partial^\rho\Big(\frac{1}{m}G^\nu+\alpha_2F^\nu+m\alpha_1A^\nu\Big)-
e\overline{\psi}\gamma_\mu\psi=0\,,
\\[3mm]\displaystyle(i\gamma^\mu D_\mu-m)\psi=0\,,\qquad \overline{\psi}(i\gamma^\mu
\overleftarrow{D}_\mu+m)=0\,.
\end{array}\end{equation}
Here, $J_\mu=e\overline{\psi}\gamma_\mu\psi$ is the current of the
spinor field, $\gamma$'s are the $3d$ gamma matrices, and $D$ is the
covariant derivative,
\begin{equation}\label{JD}
D_\mu=\partial_\mu-ie\mathcal{A}_\mu\,,\qquad
\overleftarrow{D}_\mu=\overleftarrow{\partial}_\mu+ie\mathcal{A}_\mu\,,\qquad \mathcal{A}_\mu=g_3\frac{1}{m^2}G_\mu+g_2\frac{1}{m}F_\mu+g_1A_\mu\,.
\end{equation}
The spinors $\psi,\overline{\psi}$ are Grassmann odd fields. The
spinor field $\psi$ and its conjugate $\overline{\psi}$ are
considered as independent variables. The real constants
$g_1,g_2,g_3$ are dimensionless parameters of interaction. The
parameter $e$ is a coupling constant.

In general, the interaction vertices are non-Lagrangian in the
equations (\ref{int}). The Lagrangian case corresponds to $g_1\neq
0, \, g_2=g_3=0$ in (\ref{JD}). As we shall see, the Lagrangian
model is unstable, while the stability can be retained by admitting
non-Lagrangian higher-derivative contributions to the interaction,
i.e. by $g_2\neq0, g_3\neq 0$. As we shall demonstrate in this
section, with non-Lagrangian stable interactions, the equations
(\ref{int}), (\ref{JD}) still admit constrained Hamiltonian
formulation with on-shell bounded Hamiltonian.

The consistency of interaction implies that the gauge transformation
(\ref{gt}) is complimented by the standard $U(1)$-transformation for
the spinor field
\begin{equation}\label{gtpsi}
\delta_\xi\psi=-ieg_1\psi\xi(x)\,,\qquad
\delta_\xi\overline{\psi}=ieg_1\overline{\psi}\xi(x)\,.
\end{equation}
The non-linear theory describes propagation of the gauge field $A$
coupled to the spinor $\psi$ in the gauge-invariant way.

The equations (\ref{int}) admit the second-rank conserved tensor
\begin{equation}\label{Tmnint}\begin{array}{l}\displaystyle
T_{\mu\nu}(g)=T_{\mu\nu}(\beta_1,\beta_2)+\frac{i}{4}\overline{\psi
}[\gamma_{\mu}D_{\nu}+\gamma_{\nu}D_{\mu}-
\gamma_{\mu}\overleftarrow{D}_{\nu}-\gamma_{\nu}\overleftarrow{D}_{\mu}]\psi-\frac12\eta_{\mu\nu}\overline{\psi}[(i\gamma^\rho D_\rho-m)-(i\gamma^\rho D_\rho+m)]\psi\,,
\end{array}\end{equation}
where  $T_{\mu\nu}(\beta_1,\beta_2)$ stands for the conserved tensor
 (\ref{Tmunu}) of the free theory with the parameters $\beta$ fixed
 by the interaction constant in the following way
\begin{equation}\label{bet1bet2}
\beta_1=g_1-\alpha_1g_3\,,\qquad \beta_2 = g_2-\alpha_2g_3\,.
\end{equation}
We chose the conserved tensor in the form (\ref{Tmnint}) because its
$00$-component does not involve time derivatives of the spinor
field. Once the time derivatives of the spinor filed are not
involved in $T_{00}(g)$, the conserved tensor still admits by
redefinition on-shell vanishing terms that involve the derivatives
of the vector field. The structure of this term is analogous to
(\ref{Tmunurd}), so we do not write these contributions explicitly.

Upon inclusion of interaction, the deformation is still conserved of
a single representative from the series of conserved tensors
(\ref{Tmunu}) admitted at free level. The parameters
$\beta_1,\beta_2$ in this conserved tensor are fixed by the
interaction constants by the formula (\ref{bet1bet2}).

The procedure of construction of the conserved tensor (\ref{Tmnint})
is analogous to that from \cite{KLS14}, Sec. 4.2, where the
couplings of Podolsky's electrodynamics with the spinor matter are
considered. This procedure preserves the relationship between the
conserved tensor and space-time translations. In particular,
(\ref{Tmnint}) is related to the invariance of model w.r.t. the
space-time translations if this is true in the linear approximation.
The necessary and sufficient condition to connect the conserved
tensor (\ref{Tmnint}) to the invariance of model w.r.t. the
space-time translations follows form (\ref{Delta}) and
(\ref{bet1bet2}). Substituting (\ref{bet1bet2}) into (\ref{Delta}),
we get
\begin{equation}\label{D}
g_1^2+\alpha_1g_2^2+\alpha_1^2g_3^2-\alpha_1g_1g_2+(\alpha_2^2-2\alpha_1)g_1g_3-\alpha_1\alpha_2g_2g_3\neq0\,.
\end{equation}
In what follows, we consider the interactions (\ref{int}), whose
parameters satisfy this condition. By this reason, we consider
(\ref{Tmnint}) as the energy-momentum tensor of the non-linear
theory (\ref{int}).

The $00$-component of the tensor (\ref{Tmnint}) reads
\begin{equation}\label{T00int}\begin{array}{c}\displaystyle
T_{00}(g)=T_{00}(g_1-\alpha_1g_3,g_2-\alpha_2g_3)+\frac{1}{2}\overline{\psi}[i(\gamma_i\partial_i-\gamma_i\overleftarrow{\partial}_i)+2e\gamma_i\mathcal{A}_i-2m]\psi\,,
\end{array}\end{equation} Depending on the vales of the parameters $g$,
this quantity can be bounded or unbounded from below.\footnote{With
the cubic interaction contribution, the conserved tensor
(\ref{T00int}) is no longer bounded in the strict sense. By saying
'bounded' we mean that the quadratic contribution in the conserved
quantity is bounded. The latter property is interpreted as stability
of the theory with respect to small fluctuations of initial data,
and it is not considered as obstruction to the stability of the
model. For example, the energy-momentum of spinor electrodynamics
includes cubic term.} The necessary and sufficient condition for
that follows from (\ref{b1b2}). It reads
\begin{equation}\label{gint}
    g_1-\alpha_1g_3>0\,,\qquad
    g_1^2+\alpha_1g_2^2+\alpha_1^2g_3^2-\alpha_1g_1g_2+(\alpha_2^2-2\alpha_1)g_1g_3-\alpha_1\alpha_2g_2g_3>0\,.
\end{equation}
In case of minimal interaction $g_1=1$, $g_2=g_3=0$, the equations
of motion (\ref{int}) are Lagrangian. However, the Lagrangian
non-linear theory is unstable because the canonical energy of the
model is unbounded. Once the condition (\ref{gint}) is satisfied,
the model is stable, while the field equations (\ref{int}) and
(\ref{JD}) are non-Lagrangian.

Let us bring the theory (\ref{int}) to the form of constrained
Hamiltonian dynamics. The first-order formulation for the model
(\ref{int}) is constructed in the same way as in the linear case.
The variables $A_i,F_i,G_i$ are introduced by the recipe
(\ref{varphi}) to absorb the time derivatives of $A$.For these
fields,  we get three equations of evolutionary type
\begin{equation}\label{dvarphig}\begin{array}{l}
\dot{A}_i=\partial_iA_0-\varepsilon_{ij}F_j\,,\\[3mm]
\dot{F}_i=\varepsilon_{ij}\bigr[\partial_k(\partial_kA_j-\partial_jA_k)-G_j\bigl]\,,\\[3mm]
\dot{G}_i=\varepsilon_{ij}\bigr[\partial_k(\partial_kF_j-\partial_jF_k)+m(\alpha_2G_j+\alpha_1mF_j)-J_j\bigl]\,,\\[3mm]
\end{array}\end{equation}
Obviously, the first pair of equations in this system have the same
form as in (\ref{dvarphi}) because they just define the new fields
introduced to absorb the time derivatives of the original field $A$.
The third equation represents the first-order form of the original
field equations, so it involves the interaction. With account of the
interaction, the constraint reads
\begin{equation}\label{Thetag}
\displaystyle\Theta\equiv\varepsilon_{ij}\partial_i\biggr(\frac{1}{m}G_j+\alpha_2F_j+\alpha_1mA_j\biggl)-J_0=0\,.
\end{equation}
Equations (\ref{dvarphig}), (\ref{Thetag}) are complimented by the
equations for the spinors from (\ref{int}):
\begin{equation}\label{Spinor1}\begin{array}{l}\displaystyle
\dot{\psi}=\gamma_0\{\gamma_j\partial_j+ie\gamma_i\mathcal{A}_i-ie\gamma_0\mathcal{A}_0-im\}\psi\,,\qquad
\dot{\overline{\psi}}=\overline{\psi}\{\gamma_j\partial_j-ie\gamma_i\mathcal{A}_i+ie\gamma_0\mathcal{A}_0+im\}\gamma_0\,.
\end{array}\end{equation}
These equations are of the first order from the outset. In this way,
we have the first-order formulation for the model (\ref{int}) which
includes equations (\ref{dvarphig}), (\ref{Thetag}) and
(\ref{Spinor1}).

We chose the following ansatz for the total Hamiltonian:
\begin{equation}\label{H1}\begin{array}{c}\displaystyle
H_{T}(g)=T_{00}(g)+\mathcal{A}_0\Theta\,,
\end{array}\end{equation}
where $g_1,g_2,g_3$ are the parameters. On
account of (\ref{T00int}), the Hamiltonian describes the same
conserved quantity as the $00$-component of the tensor
(\ref{Tmnint}). Substituting (\ref{H1}) into (\ref{multi-H-gen}),
(\ref{T-gen}), we arrive at the system of linear algebraic equations
defining the series of Poisson brackets:
\begin{equation}\label{PBint}\begin{array}{l}\displaystyle
\{A_i, H_T(g)
\}_{g}=\partial_iA_0-\varepsilon_{ij}F_j\,,\\[3mm]\displaystyle
\{F_i,H_T(g)\}_{g}=\varepsilon_{ij}\bigr[\partial_k(\partial_kA_j-\partial_jA_k)-G_j\bigl]\,,\\[3mm]\displaystyle
\{G_i,
H_T(g)\}_{g}=\varepsilon_{ij}\bigr[\partial_k(\partial_kF_j-\partial_jF_k)+m(\alpha_2G_j+\alpha_1mF_j)-J_j\bigl]\,,\\[3mm]\displaystyle
\{\psi,
H_T(g)\}_{g}=\gamma_0\{\gamma_j\partial_j+ie\gamma_i\mathcal{A}_i-ie\gamma_0\mathcal{A}_0-im\}\psi\,,\\[3mm]\displaystyle
\{\overline{\psi},
H_T(g)\}_{g}=\overline{\psi}\{\gamma_j\partial_j-ie\gamma_i\mathcal{A}_i+ie\gamma_0\mathcal{A}_0+im\}\gamma_0\,.
\end{array}\end{equation}
These relations should take into account the Grassmann parity of the
fields, so it is an even $Z_2$-graded Poisson bracket. In
particular, the brackets are symmetric of the spinor fields
$\psi,\overline\psi$.

Equations (\ref{PBint}) are consistent if the interaction parameters
satisfy condition (\ref{D}). The structure of the Poisson bracket,
however, depends on the relations between the interaction parameters
$g_1,g_2,g_3$. Below, we focus on the case $g_1\neq0$, while the
other cases can be treated in a similar way. The Poisson bracket,
being defined by equations (\ref{PBint}), reads
\begin{equation}\label{PBsolint}\begin{array}{l}\displaystyle
\{G_i(\vec{x}),G_j(\vec{y})\}_{g}=m^3
\frac{(\alpha_1-\alpha_2^2)g_1+\alpha_1\alpha_2g_2-\alpha_1^2g_3}
{g_1^2+\alpha_1g_2^2+\alpha_1^2g_3^2-\alpha_1g_1g_2+(\alpha_2^2-2\alpha_1)g_1g_3-\alpha_1\alpha_2g_2g_3}
\varepsilon_{ij}\delta(\vec{x}-\vec{y})\,,
\\[3mm]\displaystyle
\{F_i(\vec{x}),G_j(\vec{y})\}_{g}=m^2\frac{\alpha_2g_1-\alpha_1g_2}
{g_1^2+\alpha_1g_2^2+\alpha_1^2g_3^2-\alpha_1g_1g_2+(\alpha_2^2-2\alpha_1)g_1g_3-\alpha_1\alpha_2g_2g_3}
\varepsilon_{ij}\delta(\vec{x}-\vec{y})\,,\\[3mm]\displaystyle
\{F_i(\vec{x}),F_j(\vec{y})\}_{g}=\{A_i(\vec{x}),G_j(\vec{y})\}_{g}=
\frac{m(\alpha_1g_3-g_1)}{g_1^2+\alpha_1g_2^2+\alpha_1^2g_3^2-\alpha_1g_1g_2+(\alpha_2^2-2\alpha_1)g_1g_3-\alpha_1\alpha_2g_2g_3}\varepsilon_{ij}\delta(\vec{x}-\vec{y})\,,\\[3mm]\displaystyle
\{A_i(\vec{x}),F_j(\vec{y})\}_{g}=\frac{g_2-\alpha_2g_3}{g_1^2+\alpha_1g_2^2+\alpha_1^2g_3^2-\alpha_1g_1g_2+(\alpha_2^2-2\alpha_1)g_1g_3-\alpha_1\alpha_2g_2g_3}\varepsilon_{ij}\delta(\vec{x}-\vec{y})\,,\\[3mm]\displaystyle
\{A_i(\vec{x}),A_j(\vec{y})\}_{g}=\frac{1}{m}\frac{-\alpha_1g_3^2+\alpha_2g_2g_3+g_1g_3-g_2^2}
{g_1(g_1^2+\alpha_1g_2^2+\alpha_1^2g_3^2-\alpha_1g_1g_2+(\alpha_2^2-2\alpha_1)g_1g_3-\alpha_1\alpha_2g_2g_3)}\varepsilon_{ij}\delta(\vec{x}-\vec{y})\,.\\[3mm]
\end{array}\end{equation}
The spinor field $\psi$ and its Dirac conjugate
$\psi^\dagger=\overline{\psi}\gamma_0$ are conjugate w.r.t. to the
graded canonical bracket,
\begin{equation}\label{PBg}
\{\psi^\dagger_a(\vec{x}),\psi_b(\vec{y})\}_{g}=i\eta_{ab}\delta(\vec{x}-\vec{y})\,,\qquad
\{\psi_a(\vec{x}),\psi_b(\vec{y})\}_{g}=\{\psi^\dagger_a(\vec{x}),\psi^\dagger_b(\vec{y})\}_{g}=0\,.
\end{equation}
As is seen from these relations, the Poisson bracket is unique in
the non-linear theory (\ref{int}). No free parameters are involved
in the Poisson bracket (\ref{PBsolint}), (\ref{PBg}) besides the
coupling constants $g$.

 With no arbitrary
parameters involved in the Hamiltonian formulation, the non-linear
theory (\ref{int}) is not multi-Hamiltonian anymore, while the free
limit admits the two-parameter series of Hamiltonian formulations
(\ref{H}), (\ref{PBsol}). This means, the interaction preserves one
of possible Hamiltonian formulations admitted by the free theory.
This fact can be explained in various ways. The most simple
explanation is that upon inclusion of interaction, the deformation
of the unique entry still  conserves of the series of tensors
 (\ref{Tmunu}). The parameters of the series (\ref{Tmnint}) are fixed
by the interaction constants in the non-linear theory. It is the
sole conserved tensor which defines the unique Hamiltonian at
interacting level, while the corresponding Poisson bracket is fixed
by the Hamiltonian.

For every $g$ the Poisson bracket (\ref{PBsolint}) is a
non-degenerate tensor, so it has an inverse, being a symplectic
two-form. The latter defines the Hamiltonian action functional
\begin{equation}\label{SHamint1}\begin{array}{c}\displaystyle
S(g)=\displaystyle\int
\Big\{g_1(\alpha_1mA_i+\alpha_2F_i+\frac{1}{m}G_i)
\varepsilon_{ij}\dot{A}_j+\frac{1}{m}(g_1-\alpha_2g_2-\alpha_1g_3)
\varepsilon_{ij}F_i\dot{F}_j+\frac{g_2}{m^2}\varepsilon_{ij}G_i\dot{F}_j
+\frac{g_3}{m^3}\varepsilon_{ij}G_i\dot{G}_j-\\[3mm]\displaystyle
-H_{T}(g)\Big\}d^3x\,,
\end{array}\end{equation}
where $H_{T}(g)$ denotes the total Hamiltonian (\ref{H1}). For the minimal interaction $g_1=1,g_2=g_3=0$, we
get the standard Ostrogradski action
\begin{equation}\label{SHamint2}\begin{array}{c}\displaystyle
S(g)=\displaystyle\int
\Big\{(\alpha_1mA_i+\alpha_2F_i+\frac{1}{m}G_i)
\varepsilon_{ij}\dot{A}_j+\frac{1}{m} \varepsilon_{ij}F_i\dot{F}_j
-A_0\Theta-T_{00}(g)\Big\}d^3x\,.
\end{array}\end{equation}
For non-minimal interactions, we have the Hamiltonian action
functional (\ref{SHamint1}). This action is not canonically
equivalent to (\ref{SHamint2}). The non-minimal interactions are
consistent with the bounded Hamiltonian (\ref{H1}), while the
Ostrogradski Hamiltonian, which is associated with the minimal
interaction, is unbounded in all the instances.

In this way, we see that the higher-derivative field equations
(\ref{EL-MCS3}) are compatible with inclusion of non-minimal
explicitly covariant interactions (\ref{int}) such that the theory
still admits the Hamiltonian formalism with bounded Hamiltonian if
the model has a bounded conserved quantity at the free level.

\section{Concluding remarks}
Let us summarize and discuss the results. First, we have seen that
the third-order extension of the Chern-Simons admits a two-parameter
series of conserved tensors. If the equations (\ref{EL-MCS3})
describe unitary representations (cases (A),(B) in classification
(\ref{Cases})), the bounded conserved quantities are included in the
series. If the representations are non-unitary and/or indecomposable
(cases (C), (D), (E) in classification (\ref{Cases})), all the
conserved quantities are unbounded in the series. The series
includes the canonical energy-momentum which is unbounded in all the
cases. Second, we construct the constrained multi-Hamiltonian
formalism for the higher-derivative equations (\ref{EL-MCS3}). The
zero-zero components of the conserved tensors serve as Hamiltonians
in this formalism. The formulations with different Hamiltonians and
Poisson brackets result in the same equations, while the
formulations are not connected by canonical transformations. For the
cases with unitary representations, there are bounded Hamiltonians
in the series. The Ostrogradski Hamiltonian, being included in the
series, is unbounded. Third, we introduce explicitly the
Poincar\'e-covariant and gauge-invariant stable interactions in
higher-derivative dynamics. If the free theory has a bounded
conserved quantity, it is still conserved at interacting level.
After that, we demonstrate that the covariant and stable
higher-derivative interacting theory admits the canonical
formulation with the bounded Hamiltonian.

 \vspace{0.2cm}
 \noindent
{\bf Acknowledgments.} We thank A.A.~Sharapov for discussions on
various issues addressed in this work. The work is partially
supported by the RFBR grant 16-02-00284 and by Tomsk State
University Competitiveness Improvement Program. SLL acknowledges
support from the project 3.5204.2017/6.7 of Russian Ministry of
Science and Education.

\end{document}